\documentclass[aps,onecolumn]{revtex4}
\usepackage{amsfonts}
\usepackage{amsmath}
\usepackage{amssymb,epsf}

\begin{document}

\title{Topological charged black holes in massive gravity's rainbow and\\
their thermodynamical analysis through various approaches}
\author{S. H. Hendi$^{1,2}$, B. Eslam Panah$^{1}$ and S.
Panahiyan$^{1,3}$} \affiliation{$^1$ Physics Department and Biruni
Observatory, College of Sciences, Shiraz
University, Shiraz 71454, Iran\\
$^2$ Research Institute for Astronomy and Astrophysics of Maragha (RIAAM), P.O. Box 55134-441, Maragha, Iran\\
$^3$ Physics Department, Shahid Beheshti University, Tehran 19839,
Iran}

\begin{abstract}
Violation of Lorentz invariancy in the high energy quantum gravity
motivates one to consider an energy dependent spacetime with
massive deformation of standard general relativity. In this paper,
we take into account an energy dependent metric in the context of
a massive gravity model to obtain exact solutions. We investigate
the geometry of black hole solutions and also calculate the
conserved and thermodynamic quantities, which are fully reproduced
by the analysis performed with the standard techniques. After
examining the validity of the first law of thermodynamics, we
conduct a study regarding the effects of different parameters on
thermal stability of the solutions. In addition, we employ the
relation between cosmological constant and thermodynamical
pressure to study the possibility of phase transition.
Interestingly, we will show that for the specific configuration
considered in this paper, van der Waals like behavior is observed
for different topology. In other words, for flat and hyperbolic
horizons, similar to spherical horizon, a second order phase
transition and van der Waals like behavior are observed.
Furthermore, we use geometrical method to construct phase space
and study phase transition and bound points for these black holes.
Finally, we obtain critical values in extended phase space through
the use of a new method.
\end{abstract}

\maketitle

\section{Introduction \label{INT}}

It is arguable that Einstein gravity is an effective theory which
is valid in infrared (IR) limit while in ultraviolet (UV) regime,
it fails to produce accurate results. This shortcoming requires
modification in order to incorporate the UV regime. It is believed
that the Lorentz symmetry is an effective symmetry in IR limit of
quantum gravitational processes. Since the standard
energy-momentum dispersion relation depends on such symmetry, one
expects to regard the modified energy-momentum dispersion relation
in UV regime. Such modification motivates one to develop double
special relativity \cite{double1,double2}, in which this theory
has two upper bounds (speed of light ($c$) and the Planck energy
($E_{P}$)) \cite{double1}. In this theory, it is not possible for
a particle to achieve velocity and energy larger than the speed of
light and the Planck energy, respectively \cite{rain}.
Generalization of this doubly special relativity to curved
spacetime is gravity's rainbow \cite{rain}. On the other hand, if
one considers the gravity as an emerging phenomenon due to quantum
degrees of freedom, spacetime should be described with an energy
dependent metric. Hence, the spacetime is affected by a particle
probing it and since this particle can acquire a range of energies
($E$), a rainbow of energy is built. The gravity's rainbow can be
constructed through the use of deforming the standard
energy-momentum relation as $E^{2}f^{2}(\varepsilon
)-P^{2}g^{2}(\varepsilon )=m^{2}$, in which $\varepsilon
=E/E_{P}$, $E$ is the energy of that test particle probing the
geometry of spacetime and the functions $f^{2}(\varepsilon )$ and
$g^{2}(\varepsilon )$ are called rainbow functions. It is notable
that, in order to recover the standard energy-momentum relation in
the IR limit, the rainbow functions satisfy the following relation
\begin{equation}
\underset{\varepsilon \rightarrow 0}{\lim }f(\varepsilon )=\underset{%
\varepsilon \rightarrow 0}{\lim }g(\varepsilon )=1.
\end{equation}

Now, it is possible to define an energy dependent deformation of the metric $%
\hat{\mathbf{g}}$ with the following form \cite{rain}
\begin{equation}
\hat{\mathbf{g}}=\eta ^{\mu \nu }e_{\mu }(E)\otimes e_{\nu }(E),
\end{equation}%
where $e_{0}(E)=\frac{1}{f(\varepsilon )}\hat{e}_{0}$ and $e_{i}(E)=\frac{1}{%
g(\varepsilon )}\hat{e}_{i}$, in which the hatted quantities refer to the
energy independent frame.

On the other hand, gravity's rainbow has specific properties which were
highlighted in recent studies. At first, we point out that this theory
enjoys a modification in energy-momentum dispersion relation. Such
modification in the UV limit is examined in studies that were conducted in
discrete spacetime \cite{Hooft}, models based on string theory \cite%
{Kostelecky}, spacetime foam \cite{Amelino}, spin-network in loop quantum
gravity (LQG) \cite{Gambini}, non-commutative geometry \cite{Carroll},
Horava-Lifshitz gravity \cite{HoravaI,HoravaII} and also ghost condensation
\cite{Faizal2011}. In addition, the observational evidences confirm that
such modification could exist \cite{obs}. Second, it was pointed out that by
treatment of the horizon radius of black holes as radial coordinate in this
theory, the usual uncertainty principle stands \cite{LingLZ,LiLH}. It is
possible to translate the uncertainty principle ($\Delta p\geq 1/\Delta x$)
into a bound on the energy ($E\geq 1/\Delta x$) where $E$ can be interpreted
as the energy of a particle emitted in the Hawking radiation process. Also,
it has been shown that the uncertainty in the position of a test particle in
the vicinity of horizon should be equal to the event horizon radius (see
Refs. \cite{Adler,CavagliaDM,AmelinoAP,Ali} for more details) as $E\geq
1/\Delta x\approx 1/r_{+}$, in which $E$ is the energy of a particle near
the horizon that is bounded by the $E_{P}$ and cannot increase to arbitrary
values. Hence, this bound on the energy modifies the temperature and the
entropy of black holes in gravity's rainbow \cite{Ali}. Third, it was shown
that the black hole thermodynamics in the presence of gravity's rainbow is
modified. Such modification leads to results such as existence of remnant
for black holes \cite{Ali,rem} which is proposed to be a solution to
information paradox \cite{paradox}.

Recently, the theoretical aspects of gravity's rainbow have been
investigated in various contexts \cite%
{LeivaSV,GarattiniM,AliFK,GarattiniL,HendiF,ChangW,Santos,HendiFEP,CarvalhoLB}%
; different classes of black holes with different gauge fields have been
studied in Refs. \cite{Ali,GalanM,GimK,HendiPEMarXiv}. In addition, the
hydrostatic equilibrium equation of stars and the effects of this
generalization on neutron stars have been examined in Ref. \cite{TOV}. Also,
wormhole solutions in gravity's rainbow were obtained in Ref. \cite{wormhole}%
. Besides, the effects of rainbow functions on gravitational force are
studied in Ref. \cite{force rainbow}.

The fundamental motivation of considering gravity's rainbow comes from the
violation of Lorentz invariancy (or diffeomorphism invariancy) in the high
energy regime. In this regime, one may regard a massive deformation of
standard general relativity to obtain a Lorentz-invariant theory of massive
gravity as well. It was shown that the Lorentz-breaking mass term of
graviton leads to a physically viable ghost free model of gravity \cite%
{Rubakov}. Although the graviton in general relativity is considered as a
massless particle, there are some arguments regarding the existence of
massive gravitons. The first attempt for building such theory was done by
Fierz-Pauli \cite{Fierz} which Boulware and Deser have shown that it suffers
the ghost instability in nonlinear extension \cite{Deser}. There have been
several reports regarding the interaction effects of nonlinear theory of
massive gravity in the absence of ghost field \cite%
{Minjoon,RhamG,RhamGT,Hinterbichler}. Charged black holes in the presence of
massive gravity were investigated in Refs. \cite{CaiEGS,BabichevF} (see \cite%
{BabichevDZ,AlberteCM,KoyamaNT,NiePRD,Volkov} for more details regarding
considering massive gravity). Moreover, phase transition and entanglement
entropy of a specific massive theory were studied in Ref. \cite{Ghodrati}.
On the other hand, de Rham, Gabadadze and Tolley (dRGT) proposed another
theory of massive gravity \cite{de} without Boulware-Deser ghost \cite%
{de2,HassanR,HassanRS}. dRGT massive gravity employs a reference metric for
constructing mass terms. After that, Vegh used a singular metric for
constructing a dRGT like theory \cite{vegh}. In his theory, the graviton
behaves like a lattice in specific limits and a Drude peak was observed. It
was shown that for an arbitrary singular metric the mentioned theory is
ghost free \cite{singular} and enjoys stability which was addressed in Refs.
\cite{HassanR,HassanRS}. Different classes of exact black hole solutions in
the presence of this massive gravity and their thermodynamics, phase
transition, geometrical thermodynamics and their thermal stability have been
investigated \cite{Cai2015,HEPmass,HPEmass}. In addition, the holographic
superconductor-normal metal-superconductor Josephson junction for this
specific massive gravity has been studied and it was shown that massive
gravity has specific contributions to its properties \cite{peng hu}.

It has been proposed that a consistent quantum theory of the gravity may be
obtained through the use of black holes thermodynamics. The interpretation
of geometrical aspect of black holes as thermodynamical variable provides a
powerful viewpoint for constructing such theory. In addition, the recent
advances in gauge/gravity duality emphasize on importance of black holes
thermodynamics \cite%
{Witten1998,AharonyGMO,AharonyBJM,Lowe,JingC,HuSZ,CamanhoE,HuLN,JingPC,BazeiaLOR,KabatL,Polchinski,Klebanov,SonS}%
. On the other hand, the pioneering work of Hawking and Page, which was
based on the phase transition of asymptotically adS black holes \cite{page},
and also Witten's paper on the similar subject \cite{witten} highlighted the
importance of black holes thermodynamics. In order to study black hole
thermodynamics, we can use different approaches which are based on various
ensembles. One of the thermodynamical aspects of black holes is
investigation of thermal stability in the canonical ensemble. In the
canonical ensemble, the sign of heat capacity determines thermal
stability/instability of the black holes. In addition, the roots and
divergencies of heat capacity are denoted as bound and phase transition
point, respectively. Due to these reasons, black holes thermodynamics and
their thermal stability have been investigated in literature \cite%
{Myung2008,CarterN,KastorRT,CapelaN,HendiP,PerezRTT}.

Besides, the interpretation of cosmological constant as a thermodynamical
variable (pressure) has been recently employed in literatures. Such
consideration will lead to enriching the thermodynamical behavior of black
holes and observation of specific properties of usual thermodynamical
systems such as van der Waals like behavior \cite%
{DolanCQG2011,KubiznakM,CaiCLY,PoshtehMS,ChenLLJ,HendiV,MoL,ZouZW,XuZ,HendiPEPTEP,Zeng1512,Dehghani2016,Lan2015}%
, reentrant of phase transition \cite{reent} and existence of
triple point for black holes \cite{triple}. In addition, the
consequences of adS/CFT correspondence
\cite{KastorRT,Johnson,Dolan2014,CaceresNP}, ensemble dependency
of BTZ black holes \cite{ensemble} and two dimensional dilaton
gravity \cite{dilaton} justify the consideration of cosmological
constant not as a fixed parameter but as a thermodynamical
variable. Phase transition of different classes of black holes in
the presence of different gauge fields has been investigated by
employing such proportionality between the
cosmological constant and thermodynamical pressure \cite%
{CreightonM,GibbonsKK,XuCH,WeiCL,MoLX,ZengLL}. In order to study phase
diagrams for obtaining critical values, in Ref. \cite{int} a new method for
calculating these critical values was introduced. In this method, by using a
relation between the cosmological constant and pressure with denominator of
the heat capacity of black holes, one can obtain a relation for the
pressure. The maximum of pressure in this method is critical pressure in
which the phase transition takes place. Such method has been employed in
several papers \cite{HEPmass,HPEmass,int} and it was shown to be a
successful approach toward calculating the critical pressure and volume of
black holes in extended phase space.

Another approach for studying the thermodynamical phase transition of black
holes is through the use of a geometrical way. In this method, one can use
thermodynamical variables of the black holes to construct a thermodynamical
metric. The information regarding bound and phase transition point of the
corresponding black hole is within the behavior of Ricci scalar of the
constructed thermodynamical metric. In other words, bound and phase
transition point of black holes are presented as divergencies in the
mentioned Ricci scalar. A successful metric may cover bound and phase
transition point of heat capacity by divergencies in its Ricci scalar
without any extra (non-physical) divergency. Among various metrics that are
proposed, one can point out Weinhold \cite{WeinholdI}, Ruppeiner \cite%
{RuppeinerI}, Quevedo \cite{QuevedoII} and HPEM \cite{HPEMI,HPEMII,HPEMIII}.
These metrics have been employed to study the phase transition of black
holes in several papers \cite%
{HanC,BravettiMMA,Ma,GarciaMC,ZhangCY,BasakCNS,MoLW2016}. In Refs. \cite%
{HPEMI,HPEMII,HPEMIII}, it was shown that using Weinhold, Ruppeiner and
Quevedo metrics for specific black holes may lead to inconsistent picture.
In other words, bound and phase transition point may not be matched with
divergencies of the Ricci scalar completely, or extra (non-physical)
divergencies may arise. Interestingly, these shortcomings are avoided by
using the HPEM approach \cite{HPEMI,HPEMII,HPEMIII}. Motivated by the points
mentioned above, in this paper, we regard charged topological black holes in
the presence of massive gravity's rainbow.

\section{Basic equations and black hole solutions \label{FieldEq}}

The Lagrangian of massive gravity in the context of Einstein theory with a
minimally coupled linear $U(1)$ gauge field can be written as \cite{vegh}
\begin{equation}
\mathcal{L}= \mathcal{R}-2\Lambda -\mathcal{F}+m^{2}\sum_{i}^{4}c_{i}%
\mathcal{U}_{i}(g,f),  \label{Action}
\end{equation}
where $\mathcal{R}$ is the scalar curvature, $\Lambda $ is the cosmological
constant and $f$ is a fixed symmetric tensor. Also, $\mathcal{F}=F_{\mu \nu
}F^{\mu \nu }$ is the Maxwell invariant where $F_{\mu \nu }=\partial _{\mu
}A_{\nu }-\partial _{\nu }A_{\mu }$ is electromagnetic tensor field and $%
A_{\mu }$\ is the gauge potential. It is notable that, in Eq. (\ref{Action}%
), $c_{i}$'s are constants and $\mathcal{U}_{i}$'s are symmetric polynomials
of the eigenvalues of the $d\times d$ matrix $\mathcal{K}_{\nu }^{\mu }=%
\sqrt{g^{\mu \alpha }f_{\alpha \nu }}$ which can be written as
\begin{eqnarray*}
\mathcal{U}_{1} &=&\left[ \mathcal{K}\right], \; \mathcal{U}_{2} =\left[
\mathcal{K}\right] ^{2}-\left[ \mathcal{K}^{2}\right], \; \mathcal{U}_{3} =%
\left[ \mathcal{K}\right] ^{3}-3\left[ \mathcal{K}\right] \left[ \mathcal{K}%
^{2}\right] +2\left[ \mathcal{K}^{3}\right], \; \mathcal{U}_{4} =\left[
\mathcal{K}\right] ^{4}-6\left[ \mathcal{K}^{2}\right] \left[ \mathcal{K}%
\right] ^{2}+8\left[ \mathcal{K}^{3}\right] \left[ \mathcal{K}\right] +3%
\left[ \mathcal{K}^{2}\right] ^{2}-6\left[ \mathcal{K}^{4}\right].
\end{eqnarray*}

Taking into account Eq. (\ref{Action}) and using variational principle, one
can obtain the following field equations
\begin{equation}
G_{\mu \nu }+\Lambda g_{\mu \nu }+2\left( F_{\mu \rho }F_{\nu }^{\rho }-%
\frac{1}{4}g_{\mu \nu }\mathcal{F}\right) +m^{2}\chi _{\mu \nu } =0,
\label{Field equation}
\end{equation}
\begin{equation}
\partial _{\mu }\left( \sqrt{-g}F^{\mu \nu }\right) =0,
\label{Maxwell equation}
\end{equation}
in which $G_{\mu \nu }$ is the Einstein tensor and $\chi _{\mu \nu }$ is
\begin{eqnarray}
\chi _{\mu \nu } &=&-\frac{c_{1}}{2}\left( \mathcal{U}_{1}g_{\mu \nu }-%
\mathcal{K}_{\mu \nu }\right) -\frac{c_{2}}{2}\left( \mathcal{U}_{2}g_{\mu
\nu }-2\mathcal{U}_{1}\mathcal{K}_{\mu \nu }+2\mathcal{K}_{\mu \nu
}^{2}\right) -\frac{c_{3}}{2}(\mathcal{U}_{3}g_{\mu \nu }-3\mathcal{U}_{2}%
\mathcal{K}_{\mu \nu }  \notag \\
&+&6\mathcal{U}_{1}\mathcal{K}_{\mu \nu }^{2}-6\mathcal{K}_{\mu \nu }^{3})-%
\frac{c_{4}}{2}(\mathcal{U}_{4}g_{\mu \nu }-4\mathcal{U}_{3}\mathcal{K}_{\mu
\nu }+12\mathcal{U}_{2}\mathcal{K}_{\mu \nu }^{2}-24\mathcal{U}_{1}\mathcal{K%
}_{\mu \nu }^{3}+24\mathcal{K}_{\mu \nu }^{4}).
\end{eqnarray}

The main goal is obtaining topological static charged black holes with adS
asymptote in the massive gravity's rainbow. We consider the metric of $d$%
-dimensional spacetime with the following form
\begin{equation}
ds^{2}=-\frac{\psi \left( r\right) }{f^{2}(\varepsilon )}dt^{2}+\frac{1}{%
g^{2}(\varepsilon )}\left[ \frac{dr^{2}}{\psi \left( r\right) }%
+r^{2}h_{ij}dx_{i}dx_{j}\right] ,\ i,j=1,2,3,...,n~,  \label{Metric}
\end{equation}%
where $h_{ij}dx_{i}dx_{j}$ is a $(d-2)-$dimensional line element with
constant curvature $d_{2}d_{3}k$ and volume $V_{d_{2}}$, in which $d_{i}=d-i$%
. We should note that the constant $k$, which indicates the boundary of $%
t=constant$ and $r=constant$, can be a positive (elliptic), zero (flat) or
negative (hyperbolic) constant curvature hypersurface.

It is worth mentioning that due to generalization to gravity's rainbow, one
has to modify the reference metric as $f_{\mu \nu }=diag(0,0,\frac{%
c^{2}h_{ij}}{g^{2}(\varepsilon )})$, where $c$ is a positive constant. The
obtained reference metric also indicates that massive graviton in massive
gravity could acquire specific range of energies and the limitation of
gravity's rainbow is enforced on gravitons as well. Now, we use the modified
reference metric and obtain the $\mathcal{U}_{i}$ as $\mathcal{U}_{j}=\frac{%
c^{j}}{r^{j}}\Pi _{x=2}^{j+1}d_{x}$ \cite{Cai2015,PRLpaper}.

Inserting the gauge potential ansatz $A_{\mu }=h(r)\delta _{\mu }^{t}$ in
the Maxwell equation (\ref{Maxwell equation}) and considering the metric (%
\ref{Metric}), we obtain $h(r)=-\frac{q}{r^{d_{3}}}$ in which $q$ is an
integration constant and is related to the electric charge parameter. In
addition, one may use the definition of electromagnetic field tensor to find
its nonzero components as $F_{tr}=-F_{rt}=\frac{qd_{3}}{r^{d_{2}}}$.

Now, we are interested in obtaining the static black hole solutions for this
gravity. In order to obtain the metric function $\psi (r)$, one may use the
nonzero components of Eq. (\ref{Field equation}) which results into
\begin{equation}
\psi \left( r\right) =k-\frac{m_{0}}{r^{d_{3}}}-\frac{2\Lambda }{%
d_{1}d_{2}g^{2}(\varepsilon )}r^{2}+\frac{2d_{3}q^{2}f^{2}(\varepsilon )}{%
d_{2}r^{2d_{3}}}+\frac{m^{2}}{g^{2}(\varepsilon )}\left\{ \frac{cc_{1}}{d_{2}%
}r+c^{2}c_{2}+\frac{d_{3}c^{3}c_{3}}{r}+\frac{d_{3}d_{4}c^{4}c_{4}}{r^{2}}%
\right\} ,  \label{f(r)}
\end{equation}%
where $m_{0}$ is an integration constant which is related to the total mass
of solutions. It is notable that, the obtained metric function (\ref{f(r)}),
satisfies all components of the Eq. (\ref{Field equation}).

In order to investigate the geometrical structure of the solutions, we first
look for the essential singularity(ies). By calculating the Ricci and
Kretschmann scalars, we find
\begin{eqnarray}
\lim_{r\longrightarrow 0}R &\longrightarrow &\infty ,\text{ \ \ \ \ \ \ \ \
\ \ \ \ \ \ \ }\lim_{r\longrightarrow 0}R_{\alpha \beta \gamma \delta
}R^{\alpha \beta \gamma \delta }\longrightarrow \infty ,  \label{R0} \\
\lim_{r\longrightarrow \infty }R &=&\frac{2d}{d_{2}}\Lambda ,\text{ \ \ \ \
\ \ \ \ \ }\lim_{r\longrightarrow \infty }R_{\alpha \beta \gamma \delta
}R^{\alpha \beta \gamma \delta }=\frac{8d}{d_{1}d_{2}^{2}}\Lambda ^{2}.
\label{RRinf}
\end{eqnarray}

These scalars are finite for $r\neq 0$, and therefore, we conclude that
there is only one curvature singularity located at the origin ($r=0$). In
addition, Eq. (\ref{RRinf}) confirms that the asymptotical behavior of the
solutions is (a)dS. In order to study the geometrical effects of massive
term and gravity's rainbow on our solutions, we refer the reader to Refs.
\cite{HendiPEMarXiv,HEPmass}.

There are different models of the rainbow functions with various
motivations. In this paper, we try to present all analytical
relations with general form of rainbow functions. This general
form helps us to follow the trace of temporal ($f(\varepsilon )$)
and spacial ($g(\varepsilon )$) rainbow functions, separately. But
for plotting the diagrams and other numerical analysis we have to
choose a class of rainbow functions. For numerical calculations,
we are interested in a special class of rainbow functions which is
based on the constancy of velocity of the light
\cite{MagueijoSPRL}
\begin{equation}
f(\varepsilon )=g(\varepsilon )=\frac{1}{1+\lambda \varepsilon },
\label{RainbowFunctions}
\end{equation}%
where $\lambda $ is an arbitrary parameter. To state the matter
differently, hereafter, we keep the general form of rainbow
functions
in analytical relations and for numerical analysis, we use Eq. (\ref%
{RainbowFunctions}) as a typical form of rainbow functions.

\bigskip


\section{Thermodynamics \label{Thermo}}

Now, we should calculate the conserved and thermodynamic quantities of the
solutions to check the first law of thermodynamics. We use the definition of
Hawking temperature which is obtained through the concept of surface gravity
on the outer horizon $r_{+}$. We find
\begin{equation}
T=\frac{g(\varepsilon )}{4\pi f(\varepsilon )}\psi ^{\prime }(r)\left\vert
_{r=r_{+}}\right. =\frac{kd_{3}g(\varepsilon )}{4\pi f(\varepsilon )r_{+}}-%
\frac{r_{+}\Lambda }{2\pi d_{2}f(\varepsilon )g(\varepsilon )}-\frac{%
d_{3}^{2}q^{2}g(\varepsilon )f(\varepsilon )}{2\pi d_{2}r_{+}^{2d_{5/2}}}+%
\frac{m^{2}}{4\pi f(\varepsilon )g(\varepsilon )r_{+}^{3}}\mathcal{A},
\label{TotalTT}
\end{equation}
where $\mathcal{A}%
=d_{3}d_{4}d_{5}c_{4}c^{4}+d_{3}d_{4}c_{3}c^{3}r_{+}+d_{3}c_{2}c^{2}r_{+}^{2}+cc_{1}r_{+}^{3}
$.

Here, we conduct a study regarding the sign of temperature. This study is
necessary since the conditions regarding the positivity and negativity of
temperature put restrictions on solutions being physical or non-physical.

By taking a closer look at the temperature, one can see that charge term ($%
q^2-$term) is negative. Therefore, there is always an upper bound for the
electric charge of the black hole solutions, similar to Reissner-Nordstr\"{o}%
m and Kerr-Newman black holes. On the other hand, by considering $c_{i}>0$,
massive term ($m^2-$term) is always positive which makes the effects of this
term on the positivity of temperature. Unlike charge term, increasing energy
functions leads to decreasing the effects of massive term. We point out that
due to specific coupling of the energy functions with different terms, one
can modify the effects of these terms accordingly.

Considering the same order of magnitude for energy functions, we find that
for large values of them, charge term is dominant in the temperature. In
other words, in high energy regime, electromagnetic field has dominant
effect on the temperature of the black holes. On the other side, regarding
small values for the energy functions, one finds that second and last terms
of temperature are dominated. It is worthwhile to mention that for large
black holes (with large event horizon radius) the cosmological constant is
dominated for small values of energy functions. However, for small black
holes (with small event horizon radius), massive term is more considerable.
One may think about the possible relation between small horizon radius with
short range effective action of massive gravitons as an intermediate
particle.

In order to obtain the entropy of black holes in the Einstein gravity
context, one can employ the area law of the black holes. It is a matter of
calculation to show that entropy has the following form \cite%
{Beckenstein,Hunter1999,HawkingHP}
\begin{equation}
S=\frac{V_{d_{2}}}{4}\left( \frac{r_{+}}{g(\varepsilon )}\right) ^{d_{2}}.
\label{TotalS}
\end{equation}

The electric charge, $Q$, can be found by calculating the flux of the
electric field at infinity. We find
\begin{equation}
Q=\frac{d_{3}V_{d_{2}}f(\varepsilon )}{4\pi g^{d_{3}}(\varepsilon )}q.
\label{TotalQ}
\end{equation}

It was shown that by using the Hamiltonian approach, one can find the finite
mass $M$ of the black hole for massive gravity's rainbow as
\begin{equation}
M=\frac{d_{2}V_{d_{2}}}{16\pi f(\varepsilon )g^{d_{3}}(\varepsilon )}m_{0},
\label{TotalM}
\end{equation}%
in which by evaluating metric function on the event horizon ($\psi \left(
r=r_{+}\right) =0$), one can obtain
\begin{equation}
M=\frac{d_{2}V_{d_{2}}}{16\pi f(\varepsilon )g^{d_{3}}(\varepsilon )}\left(
kr_{+}^{d_{3}}-\frac{2r_{+}^{d_{1}}}{d_{1}d_{2}g^{2}(\varepsilon )}\Lambda +%
\frac{2d_{3}q^{2}f^{2}(\varepsilon )}{d_{2}r_{+}^{d_{3}}}+\frac{m^{2}%
\mathcal{B}r_{+}^{d_{5}}}{d_{2}g^{2}(\varepsilon )}\right) ,  \label{Mass}
\end{equation}%
where $\mathcal{B}%
=d_{2}d_{3}d_{4}c_{4}c^{4}+d_{2}d_{3}c_{3}c^{3}r_{+}+d_{2}c_{2}c^{2}r_{+}^{2}+cc_{1}r_{+}^{3}
$. It is notable that the electric potential, $U$, may be defined as the
gauge potential at the event horizon with respect to the reference
\begin{equation}
U=A_{\mu }\chi ^{\mu }\left\vert _{r\rightarrow \infty }\right. -A_{\mu
}\chi ^{\mu }\left\vert _{r\rightarrow r_{+}}\right. =\frac{q}{r_{+}^{d_{3}}}%
.  \label{TotalU}
\end{equation}

Now, we are in a position to check the validity of the first law of
thermodynamics for our solutions. It is straightforward to show that by
using thermodynamic quantities such as entropy (\ref{TotalS}), charge (\ref%
{TotalQ}) and mass (\ref{TotalM}), with the first law of black hole
thermodynamics
\begin{equation}
dM=TdS+UdQ,
\end{equation}
we can define the intensive parameters conjugate to $S$ and $Q$. These
quantities are the temperature and the electric potential
\begin{equation}
T=\left( \frac{\partial M}{\partial r_{+}}\right) _{q}\left( \frac{\partial
r_{+}}{\partial S}\right) _{q}\ \ \ \ \ \ \ \&\ \ \ \ \ \ \ \ U=\left( \frac{%
\partial M}{\partial q}\right) _{r_{+}}\left( \frac{\partial q}{\partial Q}%
\right) _{r_{+}},  \label{TU}
\end{equation}
which are coincidence with the ones calculated for the temperature and the
electric potential which are calculated in Eqs. (\ref{TotalTT}) and (\ref%
{TotalU}). As a result, the first law of thermodynamics is valid while some
of quantities are modified in the presence of massive term and rainbow
functions.

\section{Thermal stability in Canonical Ensemble \label{Stability}}

Now, our subject of interest is studying thermal stability of the solutions
in the canonical ensemble. This study is based on the behavior of heat
capacity. The negativity of the heat capacity represents an unstable state
which requires a phase transition. In the case of heat capacity, it is
argued that its divergencies are indicated with phase transition point. On
the other hand, the roots of heat capacity are the same with roots of the
temperature, so they are marking bound points which identify physical
solutions from non-physical ones. The heat capacity is given by
\begin{equation}
C_{Q}=\frac{T}{\left( \frac{\partial ^{2}M}{\partial S^{2}}\right) _{Q}}=%
\frac{T}{{\left( \frac{\partial T}{\partial S}\right) _{Q}}},  \label{CQ}
\end{equation}%
which by using Eqs. (\ref{TotalTT}), (\ref{TotalS}) and (\ref{TotalM}) leads
to
\begin{equation}
C_{Q}=\frac{\left[ \frac{2\Lambda r_{+}^{3d_{1/3}}-d_{2}m^{2}\mathcal{A}%
r_{+}^{3d_{2/3}}}{g^{d_{2}}(\varepsilon )}+\frac{d_{3}\left(
kd_{2}r_{+}^{3d_{1}}-2d_{3}q^{2}f^{2}(\varepsilon )r_{+}^{d_{-3}}\right) }{%
g^{d_{4}}(\varepsilon )}\right] r_{+}^{3}}{4d_{3}\left[ m^{2}\left(
3d_{4}d_{5}c_{4}c^{4}+2d_{4}c_{3}c^{3}r_{+}+c_{2}c^{2}r_{+}^{2}\right)
r_{+}^{2d}+g^{2}(\varepsilon )\left( k-\frac{d_{3}d_{5/2}q^{2}f^{2}(%
\varepsilon )}{3d_{2}r_{+}^{2d_{3}}}\right) +\frac{2\Lambda }{d_{2}d_{3}}%
r_{+}^{2d_{-2}}\right] }.  \label{CQQ}
\end{equation}

It is worthwhile to mention that for small value of energy functions,
cosmological constant and massive terms are dominated in the heat capacity
(the same behavior observed for the temperature). In other words, for small
energy functions, we obtain
\begin{equation}
\left. C_{Q}\right\vert _{f(\varepsilon )\sim g(\varepsilon )<<1}=\frac{%
2\Lambda r_{+}^{3d_{-2/3}}-d_{2}m^{2}\mathcal{A}r_{+}^{3d_{-1/3}}}{%
4d_{3}g^{d_{2}}(\varepsilon )\left[ m^{2}\left(
3d_{4}d_{5}c_{4}c^{4}+2d_{4}c_{3}c^{3}r_{+}+c_{2}c^{2}r_{+}^{2}\right)
r_{+}^{2d}+\frac{2\Lambda }{d_{2}d_{3}}r_{+}^{2d_{-2}}\right] },
\end{equation}%
while on the other side, like the temperature, charge term is more effective
for large values of energy functions. It is easy to obtain
\begin{equation}
\left. C_{Q}\right\vert _{f(\varepsilon )\sim g(\varepsilon )>>1}=\frac{%
-2d_{3}^{2}q^{2}f^{2}(\varepsilon )r_{+}^{d_{-6}}}{4d_{3}g^{d_{2}}(%
\varepsilon )\left( k-\frac{d_{3}d_{5/2}q^{2}f^{2}(\varepsilon )}{%
3d_{2}r_{+}^{2d_{3}}}\right) },
\end{equation}%
in which it is evident that for large values of energy functions,
all black holes with flat and hyperbolic horizons are thermally
stable, and for spherical
horizon we can obtain stable solutions provided $q^{2}>\frac{%
3d_{2}r_{+}^{2d_{3}}}{d_{3}d_{5/2}f^{2}(\varepsilon )}$.

It is not possible to obtain the root and divergencies of $C_{Q}$,
analytically. Numerical calculations show that for small values of the
horizon radius, a region of negative temperature and heat capacity exists
which is ended as these quantities meet a bound point. For specific values
of different parameters, only this bound exists which after it, system has
positive heat capacity and is in stable state. For another set of different
parameters, the phase structure is modified. In this case, the heat capacity
has one bound point and two divergencies. The stable regions are observed
between bound point and smaller divergency and also after larger divergency.
Therefore, a phase transition takes place between these two stable states.
The existence of divergencies depends on the massive parameter, energy ratio
and dimensionality. Smaller bound point and smaller divergency are
decreasing functions of massive term while the larger divergency is an
increasing function of it. As final note, it is worth mentioning that in the
absence of massive parameter and for special values of parameters, there is
no phase transition. Also regarding Eq. (\ref{RainbowFunctions}), for large
values of $\varepsilon $, hence high energy limit, similar behavior is
observed and black holes have no phase transition. This shows that the mass
of graviton and energy of particles probing the spacetime have profound
contributions into thermal stability, phase transition and thermodynamical
structure of the black holes.

\section{Extended Phase space \label{PV}}

In this section, we intend to employ the analogy between the
cosmological constant and thermodynamical pressure to study the
critical behavior of the system. It should be pointed out that the
usual relation between cosmological constant and pressure could be
modified due to these generalization. By investigation of the
energy-momentum tensor, we find that despite the modifications of
thermodynamical properties, the proportionality
between the cosmological constant and pressure remains unchanged ($P=-\frac{%
\Lambda }{8\pi }$). Now, by using Eq. (\ref{TotalTT}) and $P=-\frac{\Lambda
}{8\pi } $, one can find the following equation of state
\begin{equation}
P=\frac{d_{2}\left( 4\pi f(\varepsilon )g(\varepsilon
)Tr_{+}^{3}-kd_{3}g^{2}(\varepsilon )r_{+}^{2}-m^{2}\mathcal{A}\right) }{%
16\pi r_{+}^{4}}+\frac{2d_{3}^{2}q^{2}f^{2}(\varepsilon )g^{2}(\varepsilon
)r_{+}^{-2d_{4}}}{16\pi r_{+}^{4}}  \label{PP}
\end{equation}

The conjugating quantity of the pressure is thermodynamical volume which is
obtained by derivation of the enthalpy with respect to pressure. The
enthalpy of black holes in case of cosmological constant as pressure is
total mass. The reason is that by such consideration, the cosmological
constant is no longer a fixed parameter but a thermodynamical one.
Therefore, the phase space is extended and the mass is regarded as enthalpy.
Regarding this point, one can calculate the volume as
\begin{equation}
V=\left( \frac{\partial H}{\partial P}\right) _{S,Q}=\left( \frac{\partial M%
}{\partial P}\right) _{S,Q}=\frac{r_{+}^{d_{1}}}{d_{1}f(\varepsilon
)g^{d_{1}}(\varepsilon )}.  \label{VV}
\end{equation}

It is evident that the thermodynamical volume is modified due to the
existence of gravity's rainbow and it is a decreasing function of the energy
functions.

Due to modification in interpretation of the total mass of the black holes
in extended phase space, the Gibbs free energy is given by
\begin{equation}
G=H-TS=M-TS,  \label{G}
\end{equation}%
which by using Eqs. (\ref{TotalTT}), (\ref{TotalS}) and (\ref{TotalM}) with (%
\ref{G}), one can obtain
\begin{equation}
G=\frac{4q^{2}d_{1}d_{3}d_{5/2}f^{2}(\varepsilon )g^{d}(\varepsilon
)r_{+}^{-d_{6}} -d_{1}g^{d_{2}}(\varepsilon )r_{+}^{d_{2}}\left[
2kd_{2}^{2}g^{2}(\varepsilon )r_{+}^{2}+m^{2}\left( d_{2}\mathcal{A}+d_{1}%
\mathcal{B}\right) \right] }{16\pi d_{1}d_{2}r_{+}^{3}f(\varepsilon
)g^{2d_{3/2}}(\varepsilon )}.  \label{GG}
\end{equation}

Obtained relation for volume could be used to define a specific volume which
is related to horizon radius. This relation enables us to use horizon radius
instead of volume for studying the critical behavior of the system. In order
to obtain critical values, one can use the following properties of
inflection point in isothermal $P-r_{+}$ diagrams
\begin{equation}
\left( \frac{\partial P}{\partial r_{+}}\right) _{T}=\left( \frac{\partial
^{2}P}{\partial r_{+}^{2}}\right) _{T}=0,  \label{infel}
\end{equation}%
which for these specific class of charged black holes in the presence of
massive gravity's rainbow will lead to
\begin{equation}
d_{3}\left\{ -4d_{3}d_{5/2}q^{2}f^{2}(\varepsilon )g^{2}(\varepsilon
)r_{+}^{-2d_{4}}+kg^{2}(\varepsilon )r_{+}^{2}+m^{2}\left[
6d_{4}d_{5}c_{4}c^{4}+3d_{4}c_{3}c^{3}r_{+}+c^{2}c_{2}r_{+}^{2}\right]
\right\} =0.  \label{rc}
\end{equation}

In general, due to the contributions of massive term, it is not possible to
obtain critical horizon radius analytically. Therefore, for the simplicity
and in order to obtain critical values analytically, we exclude the effects
of $c_{3}$ and $c_{4}$. So, the critical horizon radius for arbitrary
dimensions ($d\geq 4$) is
\begin{equation}
r_{c}=\left( \frac{4d_{3}d_{5/2}q^{2}f^{2}(\varepsilon )g^{2}(\varepsilon )}{%
kg^{2}(\varepsilon )+m^{2}c^{2}c_{2}}\right) ^{\frac{1}{2d_{3}}}.
\label{rcd}
\end{equation}

Obtained relation confirms that, interestingly, one can find critical
horizon for flat, spherical and hyperbolic horizons which highlight the
effects of such configuration (charged Einsteinian black holes in the
presence of massive gravity's rainbow). It is notable that such behavior and
property was not observed for other type of black holes and it is a unique
property of these black holes. Using Eq. (\ref{infel}) with equation of
state, one can find critical temperature and pressure as
\begin{equation}
T_{c}=\frac{m^{2}cc_{1}}{4\pi f(\varepsilon )g(\varepsilon )}+\frac{d_{3}}{%
\pi f(\varepsilon )g(\varepsilon )r_{c}}\left( kg^{2}(\varepsilon
)+m^{2}c^{2}c_{2}\right) -\frac{d_{3}^{2}q^{2}f(\varepsilon )g(\varepsilon )%
}{\pi r_{c}^{2d_{5/2}}},  \label{Tcd}
\end{equation}%
\begin{equation}
P_{c}=\frac{d_{3}}{16\pi r_{c}^{2}}\left[ d_{2}\left( kg^{2}(\varepsilon
)+m^{2}c^{2}c_{2}\right) -\frac{4d_{3}d_{5/2}q^{2}f^{2}(\varepsilon
)g^{2}(\varepsilon )}{r_{c}^{2d_{2}}}\right] .  \label{Pcd}
\end{equation}

\begin{figure}[tbp]
$%
\begin{array}{ccc}
\epsfxsize=4.7cm \epsffile{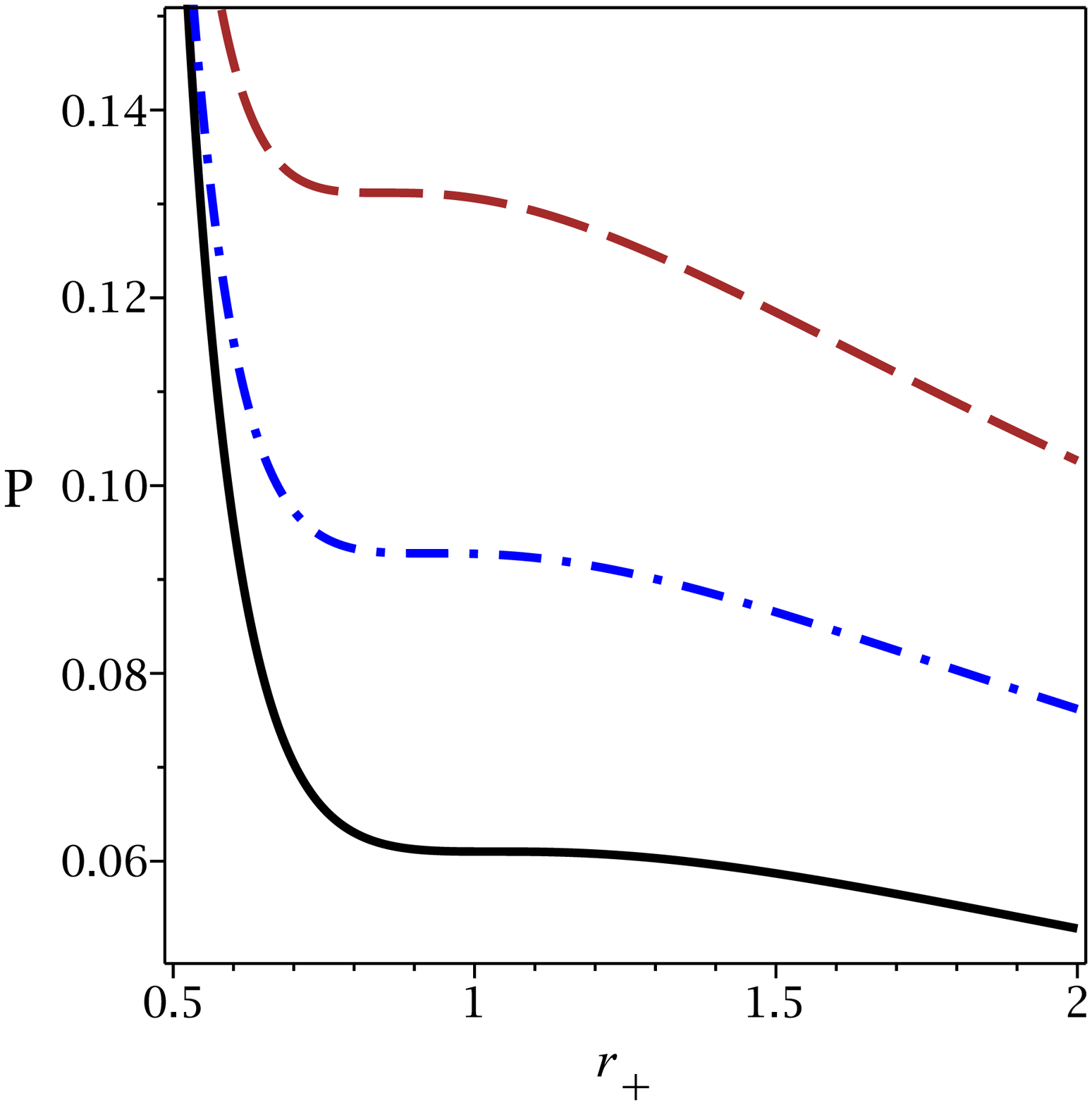} & \epsfxsize=4.7cm \epsffile{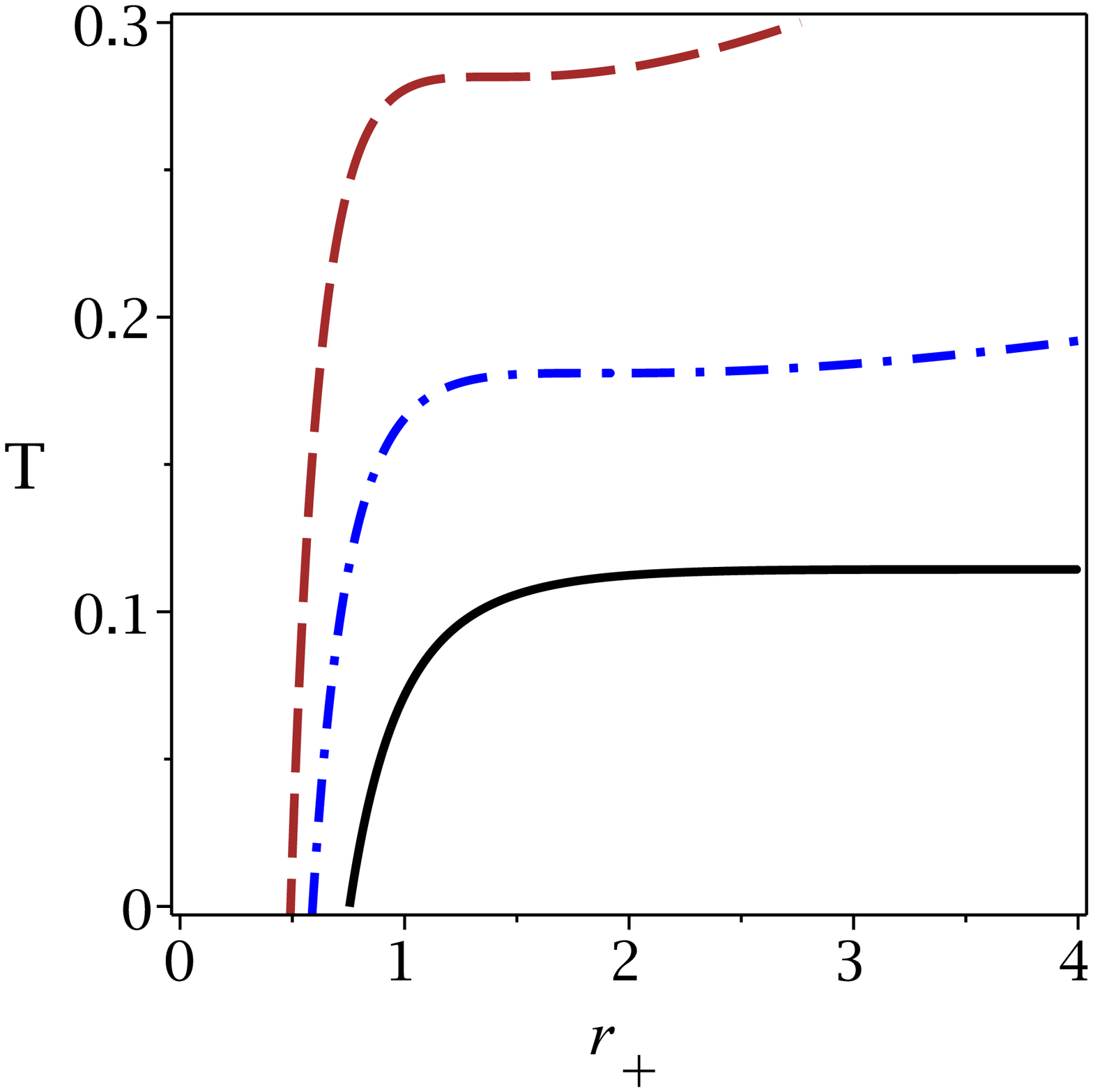}
& \epsfxsize=4.7cm \epsffile{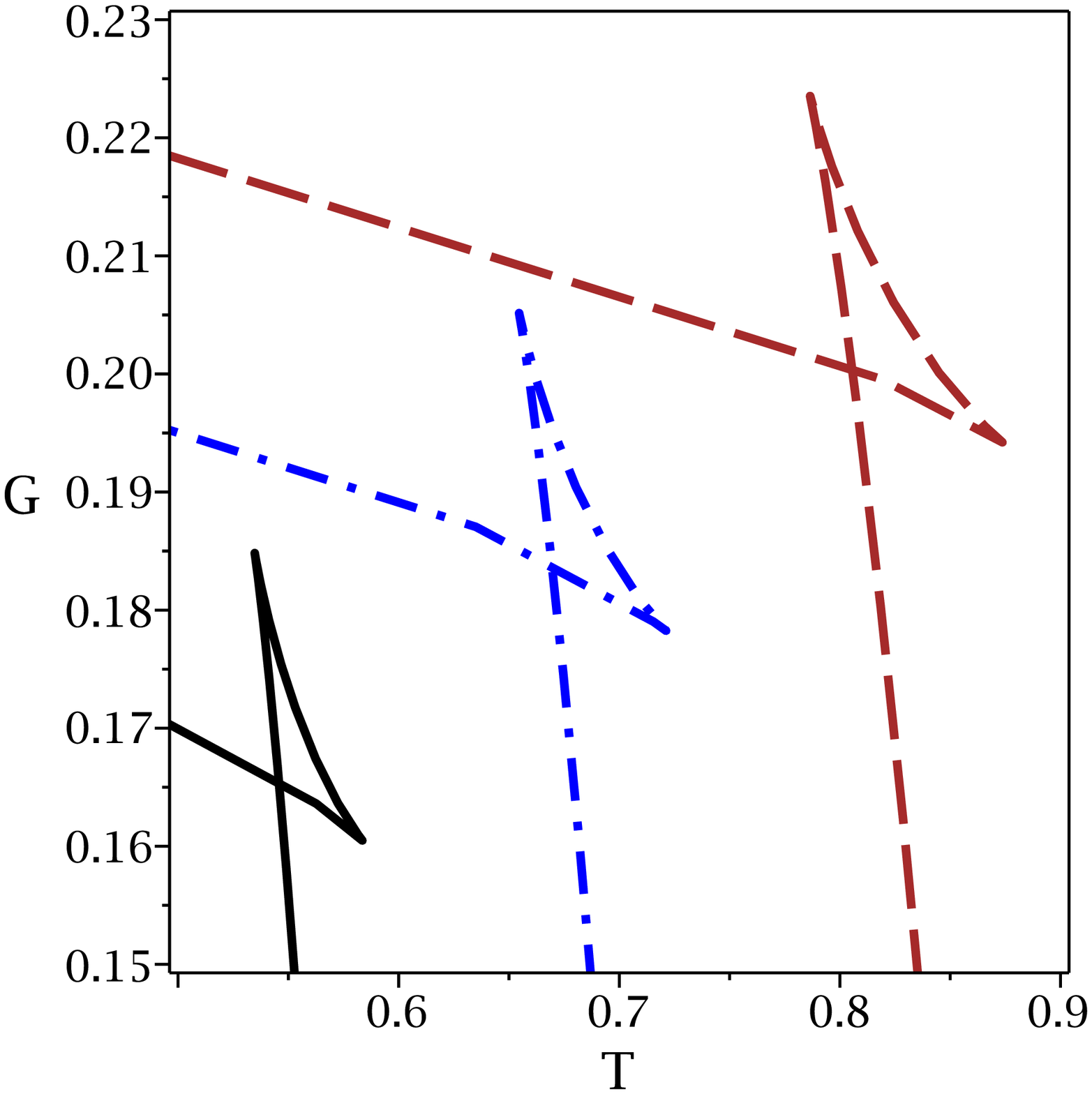}%
\end{array}
$%
\caption{ Phase diagrams for $\protect\varepsilon=0.5$, $\protect\lambda=0.3$%
, $q=m=c=c_{1}=c_{2}=1$, $c_{3}=c_{4}=0$ and $d=4$. \newline
$k=-1$ (continuous line), $k=0$ (dashed-dotted line) and $k=1$ (dashed
line). \newline
\textbf{Left panel:} $P-r_{+}$ for $T=T_{c}$; \textbf{Middle panel:} $%
T-r_{+} $ for $P=P_{c}$; \textbf{Right panel:} $G-T$ for $P=0.5P_{c}$.}
\label{FigVan}
\end{figure}

The critical horizon radius is a decreasing function of the
dimensionality and massive parameter and an increasing function of
energy functions and electric charge (if the energy functions are
considered the same). As for the critical temperature and
pressure, they are increasing functions of the dimensionality and
massive parameter and decreasing functions of energy functions and
electric charge (the energy functions are considered the same).
Using Eqs. (\ref{rcd})-(\ref{Pcd}) one can find following ratio
\begin{equation}
\frac{P_{c}r_{c}}{T_{c}}=\frac{d_{3}f(\varepsilon )g(\varepsilon )\left[
d_{2}\left( kg^{2}(\varepsilon )+m^{2}c^{2}c_{2}\right)
-4d_{3}d_{5/2}q^{2}f^{2}(\varepsilon )g^{2}(\varepsilon )r_{c}^{-2d_{2}}%
\right] }{16\left[ 2d_{3}\left( kg^{2}(\varepsilon )+m^{2}c^{2}c_{2}\right)
+m^{2}cc_{1}r_{c}-4d_{3}^{2}q^{2}f^{2}(\varepsilon )g^{2}(\varepsilon
)r_{c}^{-2d_{2}}\right] },  \label{pcrc/tcd}
\end{equation}%
where for large values of energy functions, we obtain the following energy
dependent ratio
\begin{equation}
\left. \frac{P_{c}r_{c}}{T_{c}}\right\vert _{f(\varepsilon )\sim
g(\varepsilon )>>1}=\frac{f(\varepsilon )g(\varepsilon )\left(
d_{5/2}\right) }{16}.
\end{equation}%
This ratio is an increasing function of the dimensionality and energy
functions while it is a decreasing function of the massive parameter and
electric charge. The mentioned effects are observed for spherical horizon.
The formation of swallow-tail in $G-T$ diagrams and subcritical isobar for $%
T-r_{+}$ and isothermal behavior in $P-r_{+}$ diagrams, confirm that for
hyperbolic (continuous lines of Fig. \ref{FigVan}), flat (dashed-dotted
lines of Fig. \ref{FigVan}) and spherical (dashed lines of Fig. \ref{FigVan}%
) horizons, a second order phase transition takes place. Observed
behavior for all three horizons is van der Waals like behavior.
Gibbs free energy of each phases (right panel of Fig.
\ref{FigVan}), critical pressure (left
panel of Fig. \ref{FigVan}) and temperature (middle panel of Fig. \ref%
{FigVan}) are sensitive functions of topological factor, $k$ (See
also Table \ref{tab1}.).

Next, we include the effects of $c_{3}$ and $c_{4}$. Since it is
not possible to obtain critical values analytically, we employ
numerical methods. It is evident that the largest and smallest
critical horizon radii belong to hyperbolic and spherical
horizons, respectively. On the contrary, the largest critical
temperature and pressure are obtained in spherical horizon while
the smallest ones are found for hyperbolic horizons. For the ratio
$\frac{P_{c}r_{c}}{T_{c}}$, the smallest is derived for spherical
horizon while the largest one belongs to the black holes with
hyperbolic horizons.

It is also notable that the critical values are highly sensitive to
variation of massive parameters and energy ratio in this case. In addition,
depending on being in high energy limit (large $\varepsilon$) or low energy
limit (small $\varepsilon$), the critical point of the system would be
different. This shows that consideration of the gravity's rainbow is
necessary to have a better picture regarding critical behavior of the black
holes. 
\begin{table}[tbp]
\begin{center}
\begin{tabular}{|c|c|c|}
\hline
\begin{tabular}{cccccc}
\hline\hline
\ \ $k$\ \ \  & \ \ $m$\ \  & $r_{c}$\ \ \  & \ \ \ \ $T_{c}$\ \ \ \ \ \  & $%
P_{c}$\ \ \  & $\frac{P_{c}r_{c}}{T_{c}}$ \\ \hline\hline
$-1$ & $0.9$ & $1.12051$ & $1.81518$ & $0.71931$ & $0.44403$ \\ \hline
$0$ & $0.9$ & $1.10542$ & $2.24425$ & $0.86508$ & $0.42610$ \\ \hline
$1$ & $0.9$ & $1.09189$ & $2.67889$ & $1.01467$ & $0.41357$ \\ \hline
$-1$ & $1$ & $1.06499$ & $2.61700$ & $1.08930$ & $0.44329$ \\ \hline
$0$ & $1$ & $1.05428$ & $3.06763$ & $1.25010$ & $0.42963$ \\ \hline
$1$ & $1$ & $1.04441$ & $3.52267$ & $1.41406$ & $0.41924$ \\ \hline
$-1$ & $1.1$ & $1.01777$ & $3.60572$ & $1.57104$ & $0.44345$ \\ \hline
$0$ & $1.1$ & $1.00989$ & $4.07669$ & $1.74668$ & $0.43269$ \\ \hline
$1$ & $1.1$ & $1.00250$ & $4.55123$ & $1.92499$ & $0.42402$ \\ \hline
\end{tabular}
& \ \ \ \ \ \  &
\begin{tabular}{cccccc}
\hline\hline
$k$ & $\varepsilon $ & $r_{c}$ & $T_{c}$ & $P_{c}$ & $\frac{P_{c}r_{c}}{T_{c}%
}$ \\ \hline\hline
$-1$ & $0.1$ & $1.18122$ & $1.58593$ & $0.73735$ & $0.54919$ \\ \hline
$0$ & $0.1$ & $1.16385$ & $1.99321$ & $0.90108$ & $0.52615$ \\ \hline
$1$ & $0.1$ & $1.14843$ & $2.40625$ & $1.06948$ & $0.51043$ \\ \hline
$-1$ & $0.6$ & $1.03965$ & $2.93556$ & $1.19082$ & $0.42173$ \\ \hline
$0$ & $0.6$ & $1.03007$ & $3.39697$ & $1.35094$ & $0.40965$ \\ \hline
$1$ & $0.6$ & $1.02120$ & $3.86252$ & $1.51395$ & $0.40027$ \\ \hline
$-1$ & $1.0$ & $0.94982$ & $4.50073$ & $1.65745$ & $0.34978$ \\ \hline
$0$ & $1.0$ & $0.94356$ & $5.00510$ & $1.81508$ & $0.34218$ \\ \hline
$1$ & $1.0$ & $0.93764$ & $5.51273$ & $1.97475$ & $0.33587$ \\ \hline
\end{tabular}
\\ \hline
\end{tabular}%
\end{center}
\caption{For $q=c=c_{1}=c_{2}=c_{3}=c_{4}=1$, $\protect\lambda =0.3$ and $%
d=6 $. \textbf{Left:} variation of $m$; $\protect\varepsilon =0.5$. \textbf{%
Right:} variation of $\protect\varepsilon $; $m=1$.}
\label{tab1}
\end{table}


\section{Geometrical thermodynamics}

\begin{figure}[tbp]
$%
\begin{array}{ccc}
\epsfxsize=4.7cm \epsffile{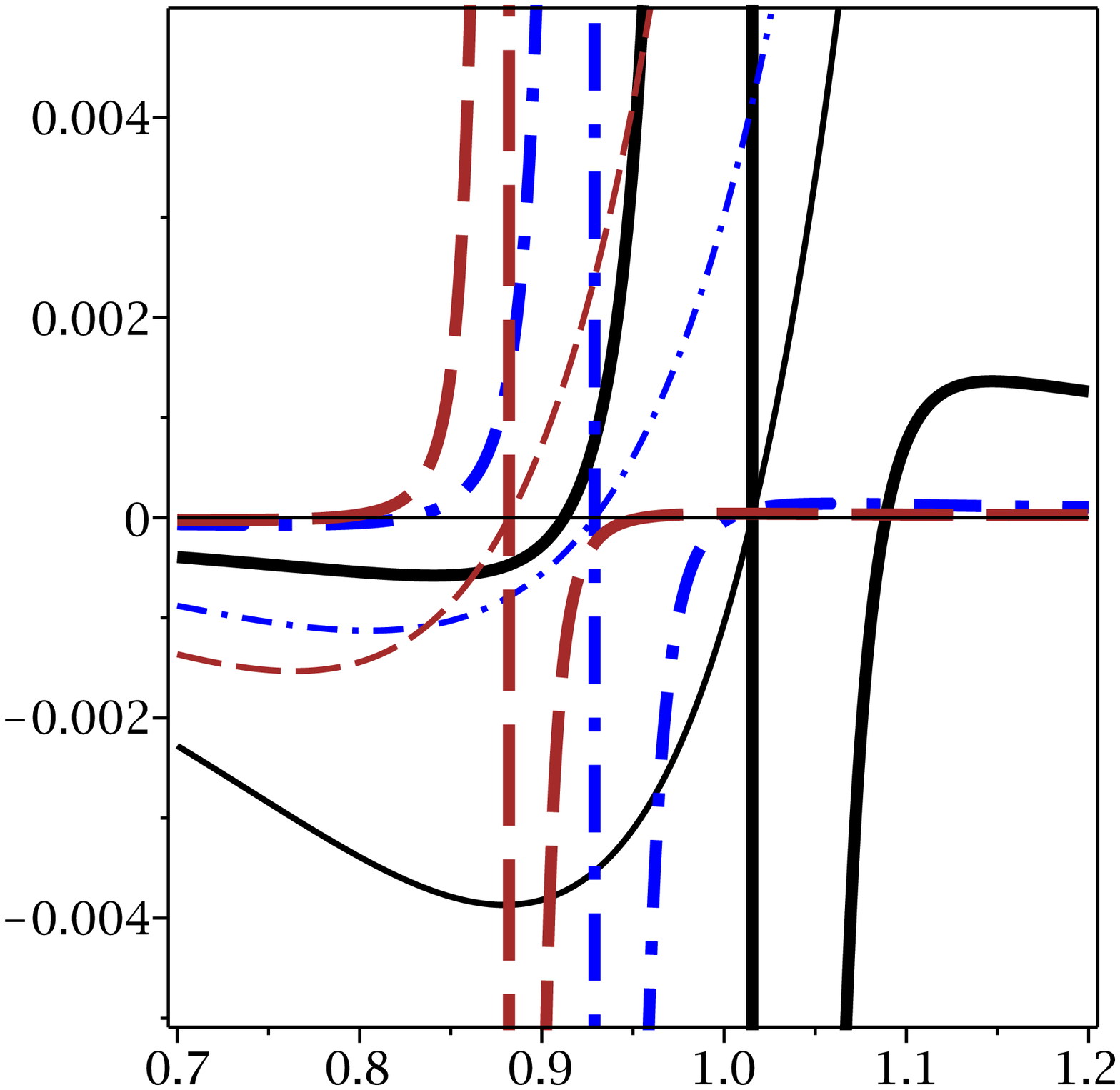} & \epsfxsize=4.7cm %
\epsffile{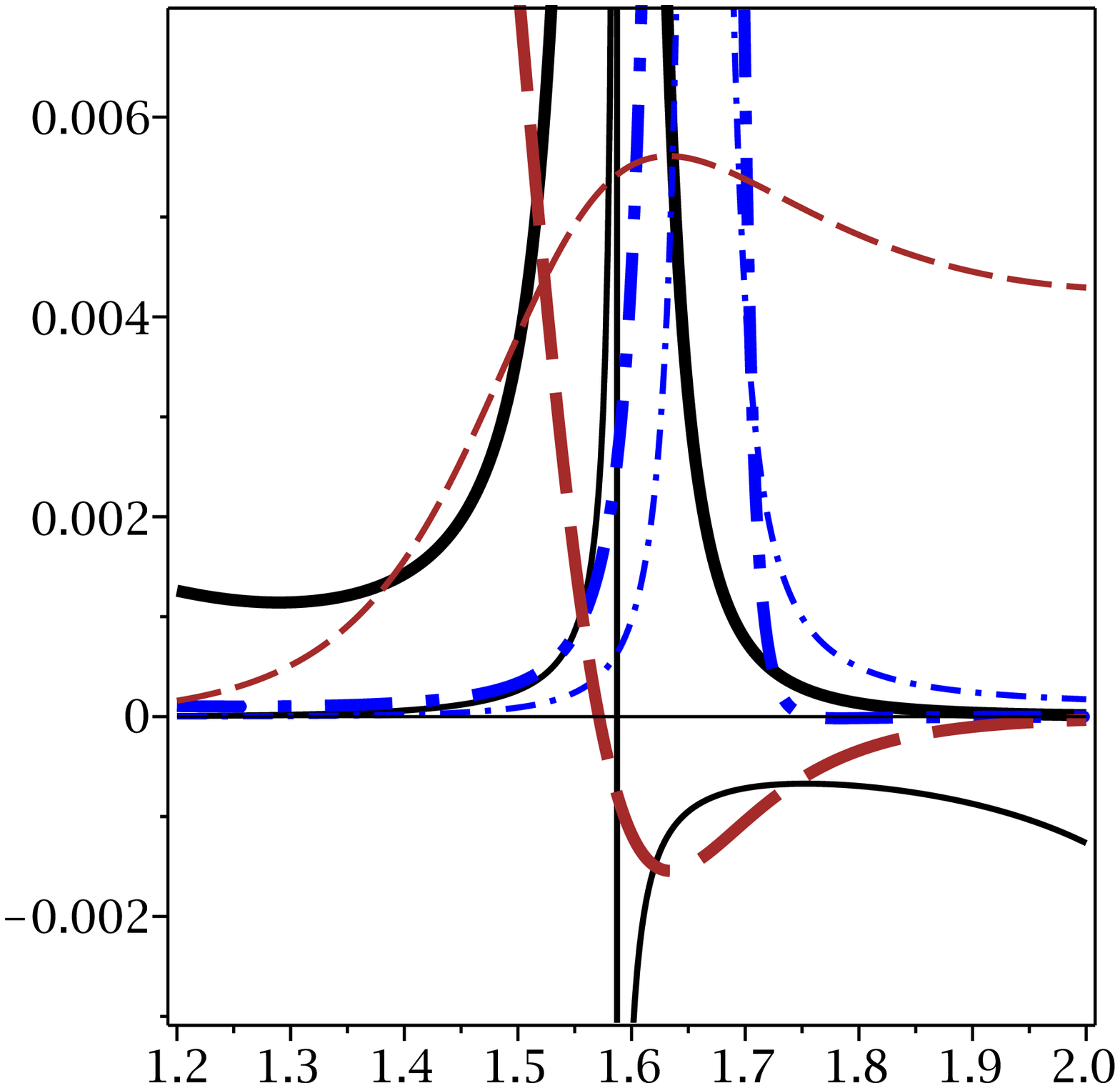} & \epsfxsize=4.7cm \epsffile{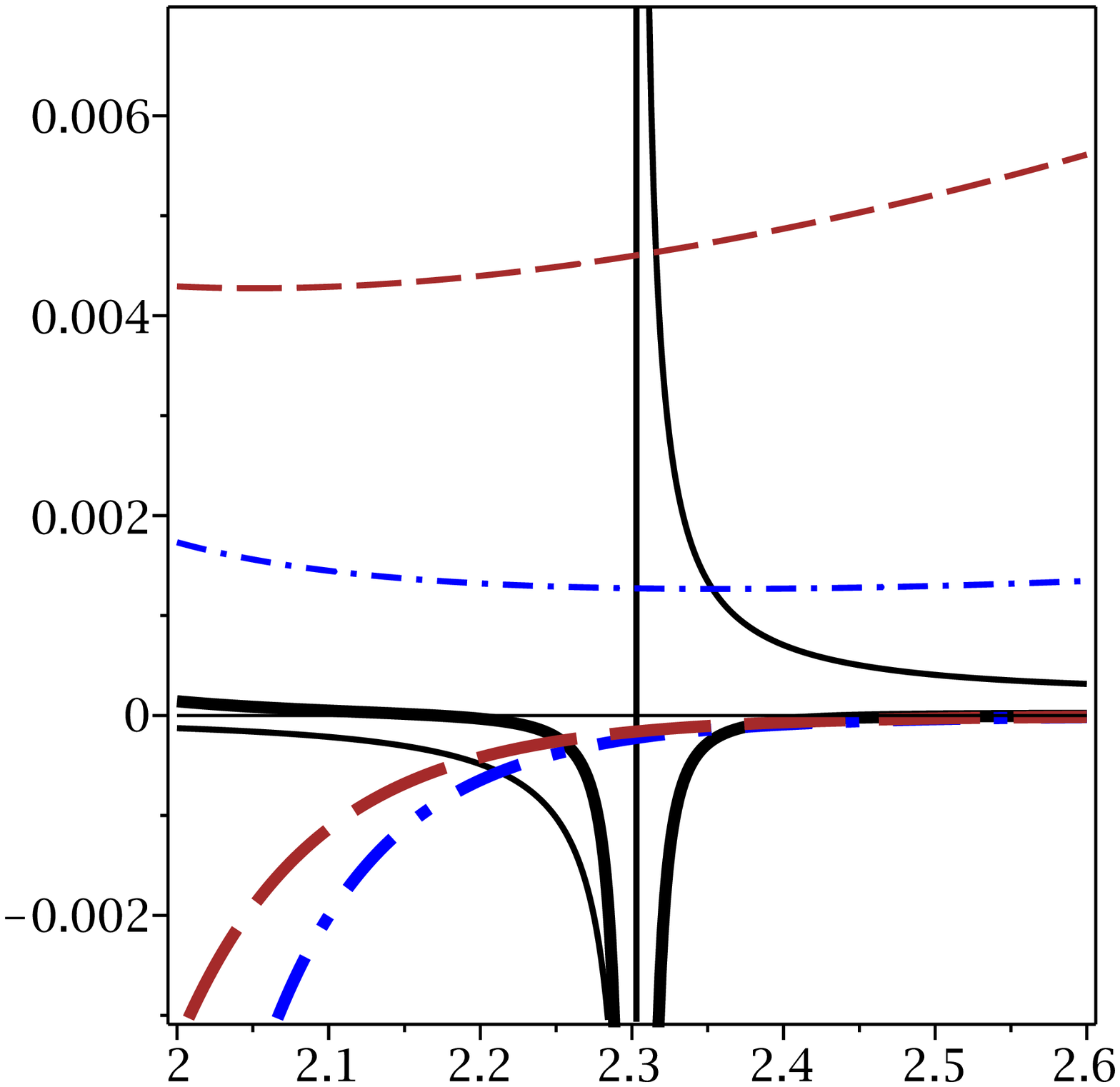}%
\end{array}
$%
\caption{For different scale: $\mathcal{R}$ (bold lines) and $C_{Q}$ versus $%
r_{+}$ for $q=1$, $c=1$, $c_{1}=c_{2}=c_{3}=c_{4}=0.1$, $k=1$, $\Lambda=-1$,
$\protect\lambda=0.5$, $m=0.3$, $d=6$, $\protect\varepsilon=0.1$ (continuous
line), $\protect\varepsilon=0.639$ (dashed-dotted line) and $\protect%
\varepsilon=1$ (dashed line).}
\label{GTD}
\end{figure}

In this section, we employ geometrical approach toward phase transition of
the black holes in the canonical ensemble. The idea is to build the phase
space of black holes through the use of thermodynamical variables. There are
several approaches with different metric constructions toward this matter
which include Weinhold \cite{WeinholdI}, Ruppeiner \cite{RuppeinerI},
Quevedo \cite{QuevedoII} and HPEM \cite{HPEMI,HPEMII,HPEMIII}. In these
approaches, the phase transition and bound points should be represented as
divergencies of the calculated Ricci scalar. In previous studies, it was
pointed out that the Ricci scalars of the Weinhold, Ruppeiner and Quevedo
metrics may lead to extra divergencies which are not matched with bound
points and phase transition \cite{HPEMI,HPEMII,HPEMIII}. Therefore, they may
result into misleading conclusions. In order to avoid this problem, another
metric was proposed before \cite{HPEMI,HPEMII,HPEMIII}. This metric contains
information which enables one to determine the type of phase transition and
distinguish divergencies related to phase transition and those correspond to
bound points. The HPEM metric has the following structure \cite%
{HPEMI,HPEMII,HPEMIII}
\begin{equation}
ds_{HPEM}^{2}=\frac{SM_{S}}{\left( \Pi _{i=2}^{n}\frac{\partial ^{2}M}{%
\partial \chi _{i}^{2}}\right) ^{3}}\left(
-M_{SS}dS^{2}+\sum_{i=2}^{n}\left( \frac{\partial ^{2}M}{\partial \chi
_{i}^{2}}\right) d\chi _{i}^{2}\right) ,  \label{HPEM}
\end{equation}%
where $M_{S}=\partial M/\partial S$, $M_{SS}=\partial ^{2}M/\partial S^{2}$
and $\chi _{i}$ ($\chi _{i}\neq S$) are extensive parameters. Here, our
solutions are static charged black holes. Therefore, the other extensive
parameter will be total electric charge. By using Eqs. (\ref{TotalS}), (\ref%
{TotalQ}) and (\ref{TotalM}) with (\ref{HPEM}), one can construct
phase space and calculate its Ricci scalar. For the economical
reasons, we will not present the calculated Ricci scalar, instead
by employing the values that are used for plotting heat capacity
diagrams, we plot some diagrams (Fig. \ref{GTD}). A simple
comparison between plotted diagrams for the heat capacity and
thermodynamical Ricci scalar shows that employed approach provides
satisfactory results. In addition, the signs of divergencies of
the Ricci scalar are different, which are depending on their
matching on the phase transition or bound points.

\begin{figure}[tbp]
$%
\begin{array}{ccc}
\epsfxsize=4.7cm \epsffile{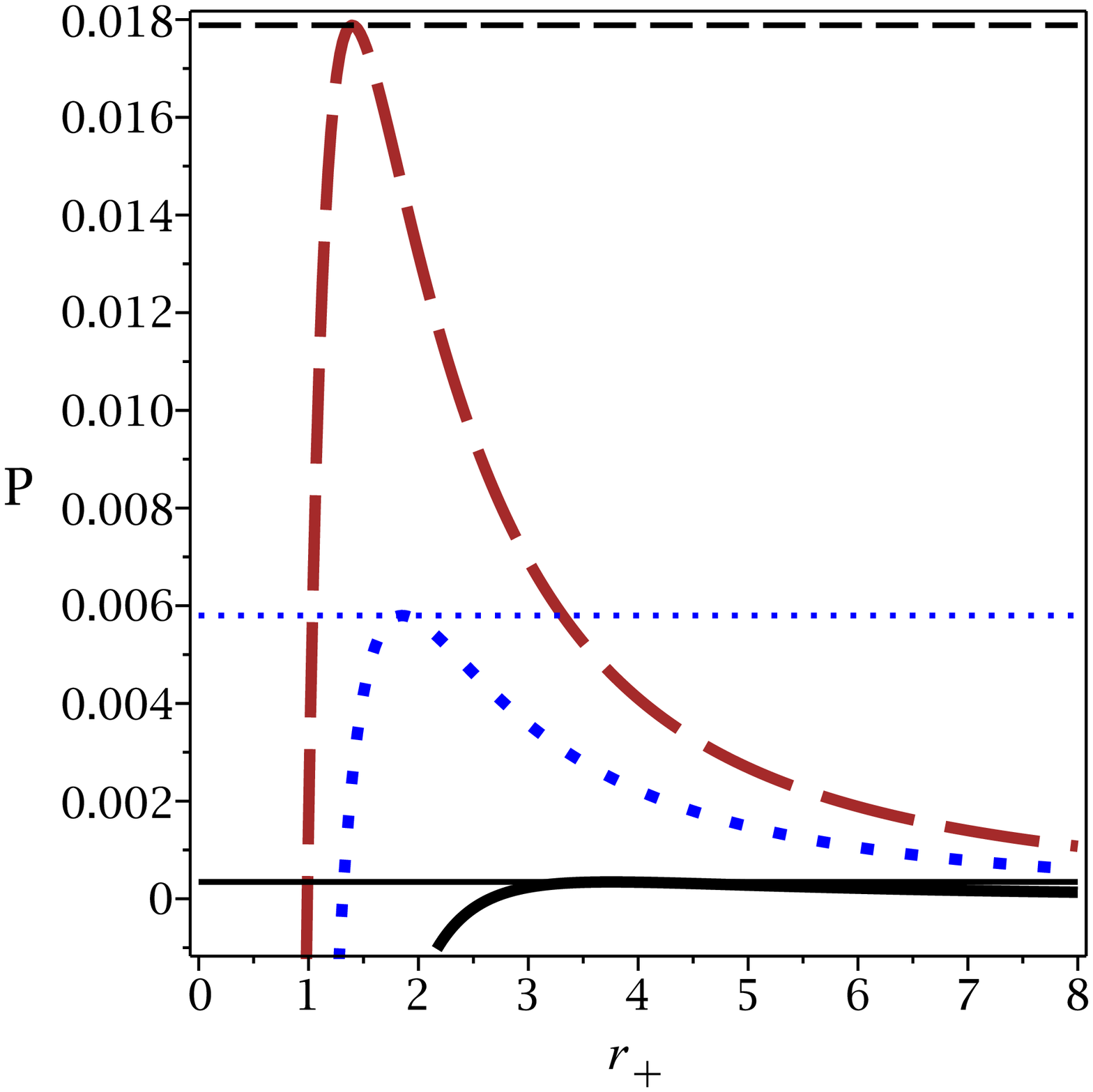} & \epsfxsize=4.7cm %
\epsffile{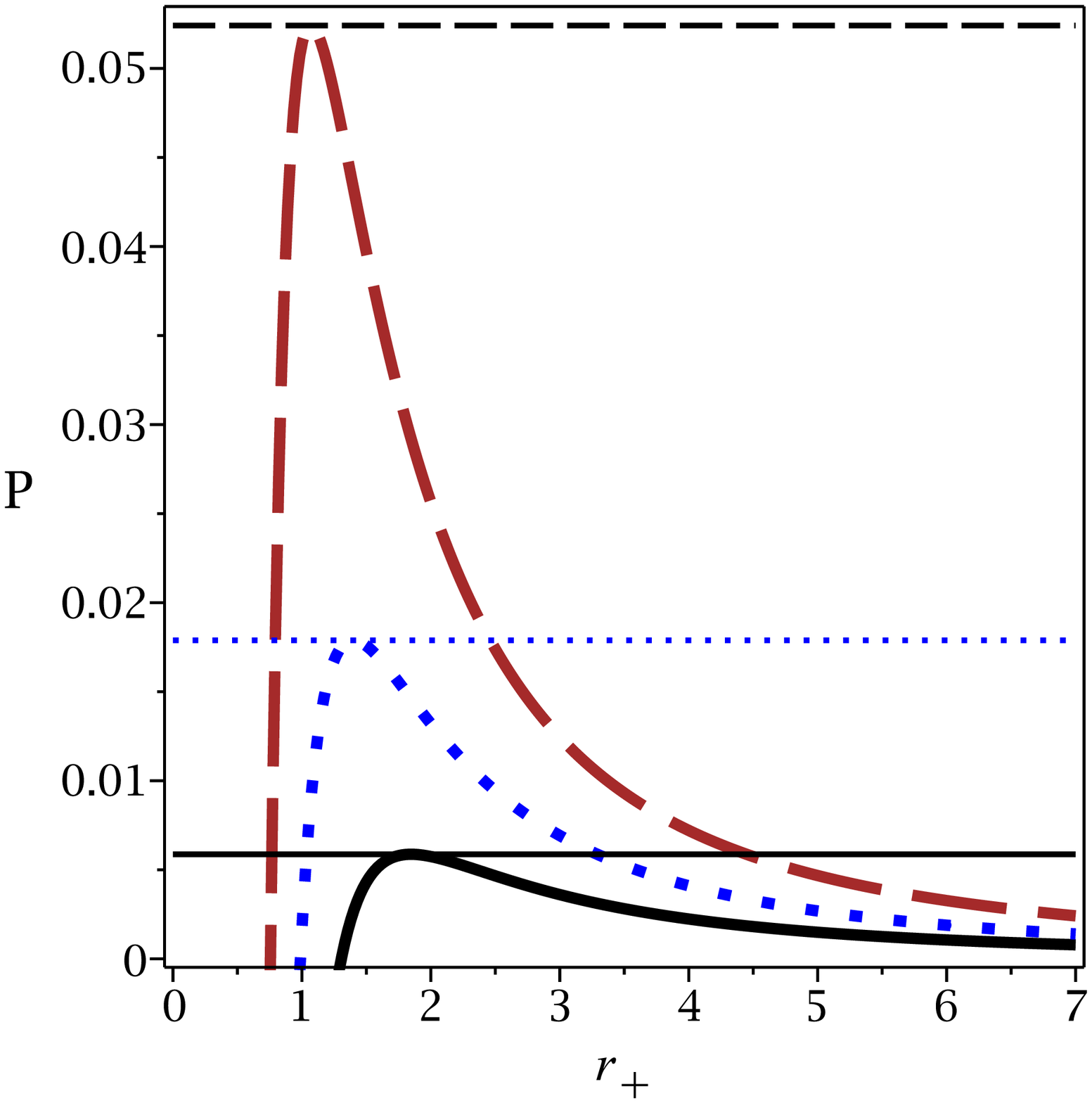} & \epsfxsize=4.7cm \epsffile{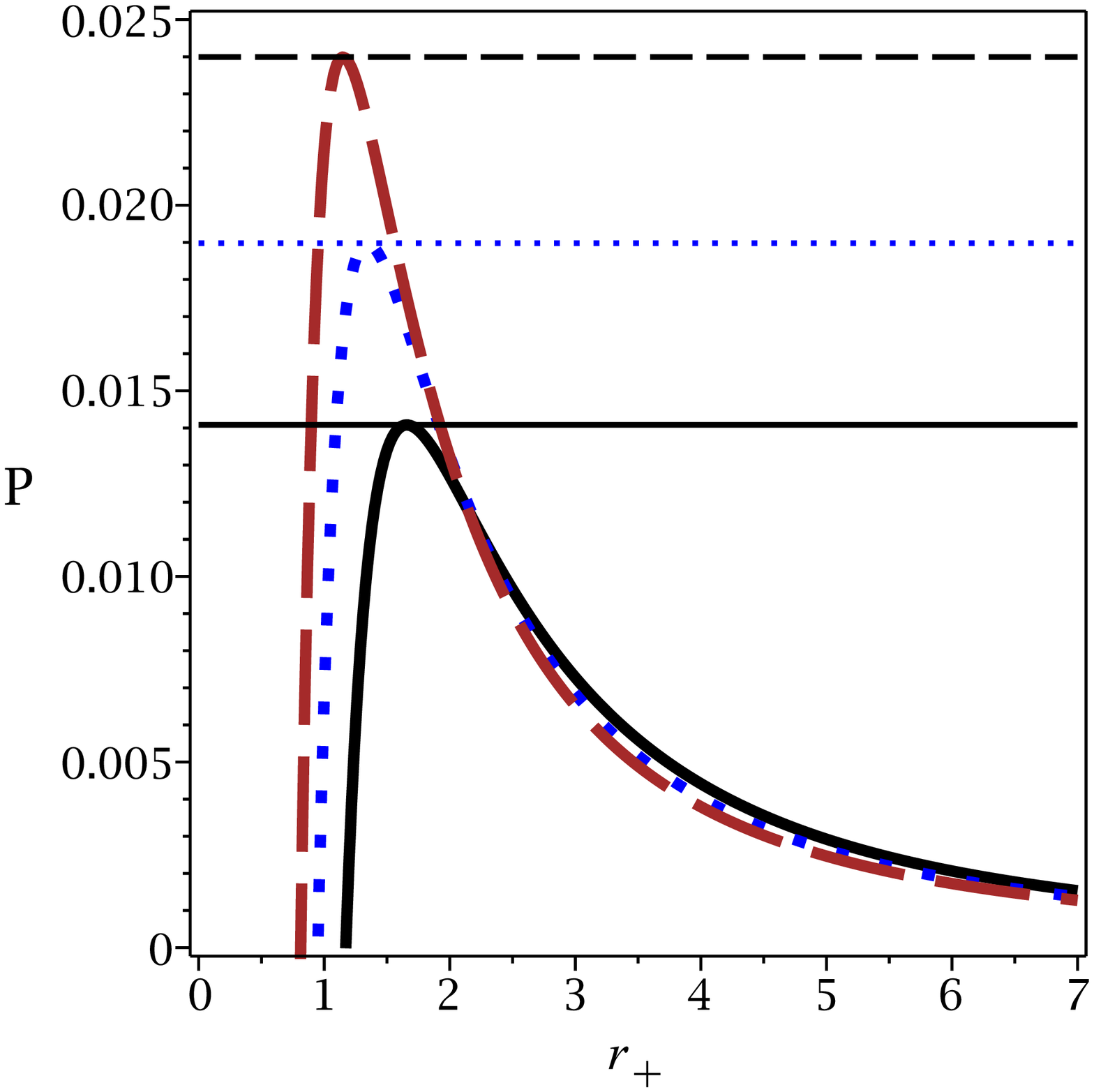}%
\end{array}
$%
\caption{$P$ versus $r_{+}$ diagrams for $\protect\lambda=0.3$, $%
q=c=c_{1}=c_{2}=1$, $c_{3}=c_{4}=0$ and $d=4$. \textbf{Left panel:} $m=1$, $%
\protect\varepsilon=0.5$, $k=-1$ (bold continuous line), $P=0.00034$
(continuous line), $k=0$ (bold dotted line), $P=0.00579$ (dotted line), $k=1$
(bold dashed line), $P=0.01788$ (dashed line). \textbf{Middle panel:} $k=1$,
$\protect\varepsilon=0.5$ $m=0.5$ (bold continuous line), $P=0.00587$
(continuous line), $m=1$ (bold dotted line), $P=0.017885$ (dotted line), $%
m=1.5$ (bold dashed line), $P=0.05240$ (dashed line). \textbf{Right panel:} $%
k=1$, $m=1$ $\protect\varepsilon=0.1$ (bold continuous line), $P=0.01408$
(continuous line), $\protect\varepsilon=0.6$ (bold dotted line), $P=0.01897$
(dotted line), $\protect\varepsilon=1$ (bold dashed line), $P=0.02399$
(dashed line).}
\label{NewMethod}
\end{figure}

\section{Maximum of pressure}

Our final study will be related to obtaining critical pressure through the
use of denominator of heat capacity. Previously, it was shown that for
finding critical pressure in extended phase space, one can use the
denominator of heat capacity \cite{HEPmass,HPEmass,int}. To do so, one
should solve the denominator of heat capacity with respect to pressure. This
method leads to an independent relation for the pressure. The existence of
maximum for pressure marks the possibility of second order phase transition.
The maximum of pressure corresponds to the critical pressure which is
observed in phase diagrams. For pressures smaller than critical pressure, a
phase transition is observed whereas for pressures larger than critical
pressure no phase transition is observed. This behavior is consistent with
what is observed in $T-V$ ($T-r_{+}$) diagrams.

Now, using Eq. (\ref{CQQ}) with $P=-\frac{\Lambda }{8\pi }$, one
can rewrite the heat capacity with dependency on thermodynamical
pressure. Solving the denominator of heat capacity with respect to
pressure, we obtain
\begin{equation}
P=\frac{d_{3}\left[ d_{2} \left( kg^{2}(\varepsilon )r_{+}^{2}+m^{2}\left[
c^{2}c_{2}r_{+}^{2}+2d_{4}c^{3}c_{3}r_{+}+3d_{4}d_{5}c^{4}c_{4}\right]%
\right) -4d_{3}d_{5/2}q^{2}f^{2}(\varepsilon )g^{2}(\varepsilon
)r_{+}^{-2d_{4}}\right] }{16\pi r_{+}^{4}}.
\end{equation}

In order to elaborate the efficiency of this method, we have
considered obtained critical values in the previous section to
plot different diagrams for the new method (Fig. \ref{NewMethod}).
It is evident that the maximum of pressure for considered values
is the same as critical pressure which could be obtained in the
extended phase space through the use of phase diagrams. Once more,
we point out that for this specific behavior a second order phase
transition is observed for non-spherical horizon. In other words,
the topological black holes irrespective of their horizon
curvatures, enjoy the existence of second order phase transition
and van der Waals like behavior in their phase diagrams. This is a
key property which was not observed before for other types of
black holes.

\section{Closing Remarks}

In this paper, we have considered massive gravity in an energy dependent
spacetime. At first, we have obtained exact charged black hole solutions and
investigated their geometrical properties. We have found that considering
massive and rainbow configurations do not change the asymptotical adS
behavior of the solutions. In addition, we have obtained the conserved and
thermodynamic quantities and their modifications are observed. Then, we have
checked the validity of the first law of thermodynamics and found that it
holds for the obtained black holes.

Next, we have investigated thermodynamical behavior of solutions
and their stability criteria. It was shown that based on the
behavior of temperature, and depending on the choices of
parameters, the physical solutions may not exist or it could be
limited to a region. Regarding thermal stability, we have shown
that adS black holes may have only one bound point or one bound
point and a phase transition. It was shown that number of
divergencies and bound points are functions of massive parameter
and gravity's rainbow. We observed that the energy of particles
probing spacetime, hence high/low energy limit affects
thermodynamical structure of the black holes, specially their
phase transition. This highlights the effects of gravity's rainbow
on the structure of solutions and emphasizes on the necessity of
such consideration.

The most interesting result of this paper was found in study that was
conducted in the context of extended phase space. Using an analytical
approach, we have shown that for these specific black holes, a second order
phase transition exists not only for spherical black holes but also for
black holes with hyperbolic and flat horizons. Such behavior was not
observed for other black holes and is a unique property of these black
holes. This shows that considering all black holes as thermodynamical
systems, the generalization to massive gravity is necessary to have a second
order phase transition and van der Waals like behavior irrespective of their
horizons structure. Such property enables one to preform specific studies in
the context of black holes with hyperbolic and flat horizons that were not
possible before due to the absence of second order phase transition and van
der Waals like behavior. In addition, the holographic phase transition and
renormalization group flow that were studied in adS/CFT context only for
spherical black holes, could be investigated through the use of these new
results. In addition, the van der Waals like behavior was also observed for
plotted $T-r_{+}$ diagram without the use of proportionality between
thermodynamical pressure and cosmological constant. In other words, the
subcritical isobar was observed.

Next, we have employed geometrical method for studying critical behavior of
the system in the context of canonical ensemble. We have indicated that
employed metric provides divergencies in its thermodynamical Ricci scalar
which coincided with bound and phase transition point of these black holes.
Our final study was conducted in regard of finding critical pressure and
horizon radius in the extended phase space through the use of denominator of
the heat capacity. We have shown that the maximum of obtained relation for
pressure using the denominator of heat capacity coincides with critical
pressure that was obtained in studying phase diagrams. Once more, it was
shown that second order phase transition is obtainable for black holes with
hyperbolic and flat horizons.

It is interesting to extend our results to higher derivative gravity in
higher dimensions as well as other gravitational models. In addition, we can
examine the causal structure of the solutions, energy conditions and their
applications with cosmological point of view. Also, one can use the
perturbation equations for such solutions to examine the well known
Gregory--Laflamme type instability \cite{GL1,GL2,GL3,GL4}. We left these
subjects for the future works.\bigskip

\begin{acknowledgements}
We would like to thank the anonymous referee for his/her valuable
comments.  We also thank Shiraz University Research Council. This
work has been supported financially by the Research Institute for
Astronomy and Astrophysics of Maragha, Iran.
\end{acknowledgements}

\end{document}